\documentclass[]{pasj01}
\draft

\begin{document} 
\Received{}
\Accepted{}

\title{Planets around the evolved stars 24 Booties and $\gamma$ Libra: A 30d-period planet and a double giant-planet system in possible 7:3 MMR}

\author{Takuya \textsc{Takarada},\altaffilmark{1,}$^{*}$ 
Bun'ei \textsc{Sato},\altaffilmark{1}
Masashi \textsc{Omiya},\altaffilmark{2}
Hiroki \textsc{Harakawa},\altaffilmark{2}
Makiko \textsc{Nagasawa},\altaffilmark{3}
Hideyuki \textsc{Izumiura},\altaffilmark{4,5}
Eiji \textsc{Kambe},\altaffilmark{4}
Yoichi \textsc{Takeda},\altaffilmark{2,5}
Michitoshi \textsc{Yoshida},\altaffilmark{6}
Yoichi \textsc{Itoh},\altaffilmark{7}
Hiroyasu \textsc{Ando},\altaffilmark{2}
Eiichiro \textsc{Kokubo},\altaffilmark{2,5} and
Shigeru \textsc{Ida}\altaffilmark{8}}
\altaffiltext{1}{Department of Earth and Planetary Sciences, School of Science, Tokyo Institute of Technology, 2-12-1 Ookayama, Meguro-ku Tokyo 152-8551}
\altaffiltext{2}{National Astronomical Observatory of Japan, 2-21-1 Osawa, Mitaka, Tokyo 181-8588, Japan}
\altaffiltext{3}{Department of Physics, Kurume University School of Medicine, 67 Asahi-machi, Kurume-city, Fukuoka 830-0011, Japan}
\altaffiltext{4}{Okayama Astrophysical Observatory, National Astronomical Observatory of Japan, Kamogata, Okayama 719-0232, Japan}
\altaffiltext{5}{The Graduate University for Advanced Studies, Shonan Village, Hayama, Kanagawa 240-0193, Japan 0000-0002-8435-2569}
\altaffiltext{6}{Hiroshima Astrophysical Science Center, Hiroshima University, Higashi-Hiroshima, Hiroshima 739-8526, Japan 0000-0002-9948-1646}
\altaffiltext{7}{Nishi-Harima Astronomical Observatory, Center for Astronomy, University of Hyogo, 407-2, Nishigaichi, Sayo, Hyogo 679-5313, Japan}
\altaffiltext{8}{Earth-Life Science Institute, Tokyo Institute of Technology, 2-12-1 Ookayama, Meguro-ku, Tokyo 152-8551, Japan}
\email{takarada@geo.titech.ac.jp}



\maketitle

\begin{abstract}
We report the detection of planets around two evolved giant stars from radial velocity measurements at Okayama Astrophysical observatory.
24 Boo (G3IV) has a mass of $0.99\,M_{\Sol}$, a radius of $10.64\,R_{\Sol}$, and a metallicity of ${\rm [Fe/H]}=-0.77$.
The star hosts one planet with a minimum mass of $0.91\,M_{\rm Jup}$ and an orbital period of $30.35\>{\rm d}$.
The planet has one of the shortest orbital periods among those ever found around evolved stars by radial-veloocity methods.
The stellar radial velocities show additional periodicity with $150\>{\rm d}$, which are probably attributed to stellar activity.
The star is one of the lowest-metallicity stars orbited by planets currently known. 
$\gamma$ Lib (K0III) is also a metal-poor giant with a mass of $1.47\,M_{\Sol}$, a radius of $11.1\,R_{\Sol}$, and ${\rm [Fe/H]}=-0.30$. 
The star hosts two planets with minimum masses of $1.02\>M_{\rm Jup}$ and $4.58\,M_{\rm Jup}$, and periods of $415\>{\rm d}$ and $964\>{\rm d}$, respectively.
The star has the second lowest metallicity among the giant stars hosting more than two planets.
Dynamical stability analysis for the $\gamma$ Lib system sets a minimum orbital inclination angle to be about $70^{\circ}$ and suggests that the planets are in 7:3 mean-motion resonance, though the current best-fitted orbits to the radial-velocity data are not totally regular.
\end{abstract}

\section{INTRODUCTION}
Over 600 planets have been detected around various types of stars by radial-velocity (RV) method.
About 400 out of them orbit around G- and K-type main-sequence stars ($0.7-1.5\>M_{\Sol}$).
These planets cover a wide range of orbital parameters, and they show correlations between occurrence rate of giant planets and properties of the host stars, such as mass \citep{Johnson2007} and metallicity (\cite{Gonzalez1997}; \cite{Fischer2005}). 
In the meantime, searching for planets around intermediate-mass stars ($>1.5\>M_{\Sol}$) on the main-sequence is difficult.
This is because early-type stars exhibit few absorption lines due to having high surface temperature and rotating rapidly (\cite{Lagrange2009}).
On the other hand, evolved intermediate-mass stars (giant stars and subgiant stars) have lower surface temperature and slow rotation, which is suit for precise RV measurements.
Several groups have performed planet searches with RV methods targeting evolved intermediate-mass stars: 
the Okayama Planet Search Program (\cite{Sato2005}) and its collaborative survey ``EAPS-Net" (\cite{Izumiura2005}), 
the BOAO K-giant survey (\cite{Han2010}), 
the Lick G- and K-giant survey (\cite{Frink2001}; \cite{Hekker2006}), 
the ESO planet search program \citep{Setiawan2003}, 
the Tautenburg Observatory Planet Search (\cite{Hatzes2005}; \cite{Dollinger2007}), 
the Penn State Tor${\rm \acute{u}}$n Planet Search \citep{Niedzielski2007} and its RV follow-up program Tracking Advanced Planetary Systems \citep{Niedzielski2015},
the survey ``Retired A Stars and Their Companions" \citep{Johnson2006},
the Pan-Pacific Planet Search \citep{Wittenmyer2011},
and the EXoPlanet aRound Evolved StarS project \citep{Jones2011}.
From these programs, about 150 planets have been discovered around evolved stars with surface gravity of $\log g<4$ so far.
Although the sample is still small in number compared to that around main-sequence stars, some statistical properties, which are not common for planets around main-sequence stars, are shown.

A lack of short-period planets is a well-known property of giant planets around giant stars.
Although short-period planets should be more easily detected than long-period ones especially around giant stars that show large intrinsic RV variations, most of the planets around giant stars reside at farther than $0.5\>$au from the central stars \citep{Jones2014}.
Some theoretical studies showed that stellar evolution, that is expansion of central stars, could induce the lack of short-period planets (e.g., \cite{Villaver2009}; \cite{Kunitomo2011}).
This would be the case with planets around low-mass giant stars ($\leq1.5\,M_{\Sol}$), since there are a large number of short-period planets detected around solar-mass main-sequence stars, which are progenitors of low-mass giant stars.
With regard to high-mass giant stars ($>1.5\,M_{\Sol}$), several studies predicted that short-period planets were hardly formed around high-mass stars by nature owing to shorter timescale of protoplanetary disk depletion than that of planetary migration (e.g., \cite{Burkert2007}; \cite{Kennedy2009}; \cite{Currie2009}).
A small number of short-period planets have been found around evolved stars by RV method.
For example, TYC 3667-1280-1 ($M=1.87\pm0.17\,M_{\Sol}$, $R=6.26\pm0.86\,R_{\Sol}$, $\log g=3.11\,\pm\,0.09$; \cite{Niedzielski2016}) hosts a planet with a semimajor axis $a=0.21\,\pm\,0.01\>{\rm au}$ and HD 102956 ($M=1.68\,\pm\,0.11\,M_{\Sol}$, $R=4.4\,\pm\,0.1\,R_{\Sol}$, $\log g=3.5\,\pm\,0.06$; \cite{Johnson2010}) hosts a planet with a semimajor axis $a=0.081\,\pm\,0.002\>{\rm au}$.
Recently, several short-period transiting planets have been confirmed around evolved stars by Kepler space telescope (\cite{Lillo-Box2014a}, \yearcite{Lillo-Box2014b}; \cite{Barclay2015}; \cite{Sato2015}; \cite{VanEylen2016}; \cite{Grunblatt2016}; \cite{Jones2017arXiv}).

A role of metallicity in planetary formation and evolution is not well investigated for evolved stars.
While \citet{Reffert2015}, \citet{Jones2016} and \citet{Wittenmyer2017} showed that metal-rich giant stars preferentially host giant planets, \citet{Jofre2015} showed that there is no difference in metallicity between giant stars with and without planets.
\citet{Jofre2015} also showed that the metallicity of evolved stars hosting multiple-planets is slightly enhanced compared with evolved stars hosting single-planet, which was suggested for main-sequence stars (\cite{Wright2009}).
As for low-mass giant stars, host star's metallicity may also have a relation with the orbital configuration.
\citet{Dawson2013} showed a lack of hot Jupiters ($P<10\>$days) among Kepler giant planets orbiting metal-poor (${\rm [Fe/H]}<0.0$) stars with stellar temperature of $4500<T<6500\>{\rm K}$ and $\log g>4$.
Since low-mass giant stars have a tendency to have [Fe/H] values less than zero (e.g., \cite{Maldonado2013}; \cite{Jofre2015}), the lack of hot Jupiters seen in metal-poor dwarfs may also be the case with low-mass metal-poor giant stars.
However, the number of detected planets around evolved stars are small, and thus these suggestions are not conclusive at this stage.

Multiple-planet systems are of great importance to understand the formation and evolution of planetary systems.
Until now, 21 multiple-planet systems\footnote{NASA Exoplanet Archive} have been confirmed around evolved stars with $\log g<4.0$.
Among them, some systems are considered to be in mean-motion resonances (MMR).
For example, 24 Sex b and c (\cite{Johnson2011}) and $\eta$ Cet b and c (\cite{Trifonov2014}) are in 2:1 MMR, HD 60532 b and c are in 3:1 MMR \citep{Desort2008}, HD 200964 b and c (\cite{Johnson2011}) and HD 5319 b and c (\cite{Giguere2015}) are in 4:3 MMR, and HD 33844 b and c (\cite{Wittenmyer2016}) are in 5:3 MMR. 
HD 47366 b and c (\cite{Sato2016}) and BD+20 2457 b and c (\cite{Niedzielski2009}), which are near 2:1 and 3:2 MMR, respectively, are of interest.
The dynamical stability analysis for the systems revealed the best-fit Keplerian orbits are unstable, while a retrograde configuration will retain both systems stable (\cite{Horner2014}; \cite{Sato2016}).
\citet{Sato2016} also showed that most of the pairs of the giant planets with period ratios smaller than two are found around evolved stars.

In the Okayama Planet Search Program, we have been observing about 300 G, K giant stars for about 15 years by RV method.
We have found 25 planets in this program so far.
In this paper, we report newly detected planets around two evolved stars, 24 Boo and $\gamma$ Lib.
Stellar properties and our observations are given in section 2 and 3, respectively.
In section 4, we perform orbital analysis, and also perform Ca\,\emissiontype{II} H line examination and bisector analysis to evaluate stellar activity.
We show these results in section 5, and devote section 6 to discussion including dynamical analysis for $\gamma$ Lib system.
\section{STELLAR PROPERTIES} 
24 Boo and $\gamma$ Lib are listed in the Hipparcos catalog as a G3IV and a K0III star with the apparent V-band magnitudes $V=5.58$ and $3.91$, respectively \citep{ESA1997}. 
The Hipparcos parallax of $10.00\,\pm\,0.25\>$mas and $19.99\,\pm\,0.16\>$mas \citep{van2007} give their distance of $100.0\,\pm\,2.5\>$pc and $50.02\,\pm\,0.40\>$pc, respectively.
24 Boo is a high proper-motion star and is considered to belong to thick-disk population \citep{Takeda2008}.
The stellar properties of 24 Boo and $\gamma$ Lib were derived by \citet{Takeda2008}.
They determined stellar atmospheric parameters (effective temperature, $T_{\rm eff}$; surface gravity, $\log g$; micro-turbulent velocity, $v_t$; and Fe abundance, [Fe/H]) spectroscopically by measuring the equivalent widths of Fe\,\emissiontype{I} and Fe\,\emissiontype{II} lines.
They also determined projected rotational velocity, $v\sin i$, and macro-turbulent velocity with the automatic spectrum-fitting technique \citep{Takeda1995}.
Details of the procedure are described in \citet{Takeda2002} and \citet{Takeda2008}, and the parameters they derived for 24 Boo and $\gamma$ Lib are summarized in Table~\ref{tab1}.
In the table, we define the surface gravity derived by \citet{Takeda2008} as $\log g_{\rm sp}$ and the surface gravity derived in this work (see the following paragraph) as $\log g_{\rm iso}$.

\citet{Takeda2008} estimated stellar mass for 24 Boo and $\gamma$ Lib to be $1.92\,M_{\Sol}$ and $2.15\,M_{\Sol}$, respectively, with use of their determined luminosity, $T_{\rm eff}$, and theoretical evolutionary tracks of \citet{Leje2001}. 
However, \citet{Takeda2015} and \citet{Takeda2016} reported that stellar masses derived in \citet{Takeda2008} tend to be overestimated by a factor of 2 especially for giant stars located near clump region in the HR diagram due to their use of the coarse evolutionary tracks of low-resolution parameter grid and the lack of ``He flash" red-clump tracks for lower mass stars ($M < 2\,M_{\Sol}$).
They showed the overestimation could be mitigated with use of recent theoretical isochrons computed based on the PARSEC code \citep{Bressan2012} with fine parameter grid.
Therefore we re-estimated stellar mass together with radius, age, and $\log g$, for 24 Boo and $\gamma$ Lib in this paper.

\begin{figure}
\begin{center}
\FigureFile(80mm,50mm){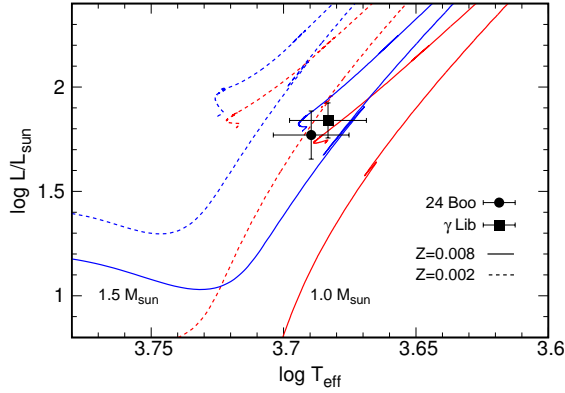}
\end{center}
\caption{The HR diagram with 24 Boo and $\gamma$ Lib. 
Some evolutionary tracks from \citet{Bressan2012} for masses of $1.0\,M_{\Sol}$ (red) and $1.5\,M_{\Sol}$ (blue) are also shown. 
The solid line and dotted line correspond to ${\rm Z}=0.008$ (${\rm [Fe/H]} = -0.29$) and ${\rm Z}=0.002$ (${\rm [Fe/H]} = -0.89$), respectively.
The error bar for effective temperature corresponds to a realistic uncertainty, $70\>{\rm K}$.
} 
\label{re_HR}
\end{figure}

For this purpose, we adopted a Bayesian estimate method with theoretical isochrones described in \citet{daSilva2006}.
We derived a posterior probability for each stellar parameter.
The posterior probability that an observed star belongs to a small section of an isochrone, $P^{12}$, is given by 
\begin{equation}
\label{eq_par}
P^{12}(x)\propto \int^{M_2}_{M_1}\phi(M)dM\ \times \exp \left[-\frac{1}{2}\sum^{N}_{i}\left(\frac{\theta_{i,{\rm obs}}-\theta_{i}'}{\sigma_{\theta_{i,{\rm obs}}}}\right)^2\right],
\end{equation}
where $M_1$ and $M_2$ are initial stellar mass at the section of the isochrone, $\phi(M)$ is the initial mass function which we assumed to be $\phi(M)=M^{-2.35}$ \citep{Salpeter1955}, $\theta_{i,{\rm obs}}$ is the observed value for the i-th stellar parameter, $\theta_{i}'$ is its theoretical value at the section of the isochrone, $\sigma_{\theta_{i,{\rm obs}}}$ is the observational error of $\theta_{i,{\rm obs}}$, and $x$ is the stellar parameter we want to estimate.
In the right side of equation~(\ref{eq_par}), the first term represents the prior probability, which is equivalent to the number of stars existing at the section of the isochrone, and the second term represents the likelihood, which is the probability that we would have the observed value, $\theta_{i,{\rm obs}}$, given the theoretical value, $\theta_{i}'$, under the assumption that the observational errors have Gaussian distributions.
For $\theta_{i}$, we adopt three parameters; absolute magnitude, effective temperature and metallicity. 
Thus $N$ in equation~(\ref{eq_par}) is $3$ in our case.
The uncertainties of stellar atmospheric parameters derived in \citet{Takeda2008} are intrinsic statistical errors and realistic uncertainties should be larger by a factor of $\sim2$.
This is because stellar atmospheric parameters are susceptible to the changes in the equivalent widths of Fe\,\emissiontype{I} and Fe\,\emissiontype{II} lines \citep{Takeda2008}.
Therefore, we here conservatively fixed the observational errors of absolute magnitude, effective temperature and metallicity to realistic values following \citet{daSilva2006}, which correspond to $0.1\>{\rm mag}$, $70\>{\rm K}$ and $0.1\>{\rm dex}$, respectively.
The stellar isochrons were taken through the CMD web interface\footnote{http://stev.oapd.inaf.it/cgi-bin/cmd} \citep{Bressan2012} which are computed based on the PARSEC code and include red-clump tracks for lower-mass stars.
We make a probability distribution function (PDF) by summing up the posterior probabilities along the isochrone, and integrating them over a range of metallicities and ages. 
We adopt a flat distribution of ages in the integration.
Then we obtained the mode value of the PDF as the most probable stellar parameters and the range between $15.87$\% and $84.13$\% of the PDF as the $1\,\sigma$ uncertainty.
The stellar parameters thus derived are listed in Table~\ref{tab1}.
We found that the stellar masses are derived to be smaller by a factor of $\sim2$ compared to Takeda et al.'s values, while $\log g$ and stellar radius agree with their values within $2\,\sigma$.
Figure~\ref{re_HR} shows the position of 24 Boo and $\gamma$ Lib in the HR diagram.
Some evolutionary tracks \citep{Bressan2012} with a similar stellar mass and metallicity to those of 24 Boo and $\gamma$ Lib are also shown.
We can see that 24 Boo is ascending the red giant branch. 
On the one hand, we can not determine the evolutionary status of $\gamma$ Lib from the HR diagram.

\begin{figure}
\begin{center}
\includegraphics[width=10cm]{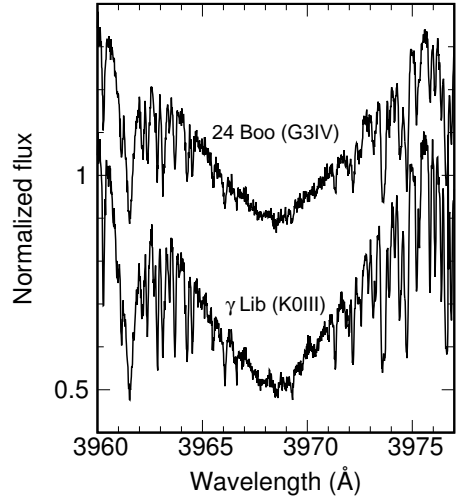}
\end{center}
\caption{The normalized spectra near Ca\,\emissiontype{II} H line for 24 Boo (top) and $\gamma$ Lib (bottom). An offset of $0.3$ is added to the spectrum of 24 Boo for clarity.}
\label{re_cahk_127_138}
\end{figure}

Both stars are known to be stable in photometry to a level of $\sigma_{\rm HIP}<0.008\ {\rm mag}$ \citep{ESA1997}.
They are also chromospherically inactive since we see no significant reversal emission in the line core of Ca\,\emissiontype{II} H, as shown in Figure~\ref{re_cahk_127_138}.
As for 24 Boo, \citet{Howard2010} showed that the star has an activity indices $\log R'_{\rm HK}=-5.19$, which supports 24 Boo is chromospherically inactive.
\begin{table}
\tbl{Stellar parameters.\footnotemark[$*$]}{%
 \begin{tabular}{lrr}
 \hline\hline
 						& 24 Boo 						& $\gamma$ Lib\\ \hline
Spectral type 				& G3IV 						& K0III \\
$V$ 				& $5.58$ 				& $3.91$ \\
$B-V$ 				& $0.864\pm0.004$ 				& $1.007\pm0.003$ \\
$M_V$ 				& $+0.69^{(a)}$ 				& $+0.45^{(a)}$ \\
$L\ (L_{\odot})$ 		& $59.45\pm6.85^{(a)}$ 				& $70.79\pm5.95^{(a)}$ \\
$T_{\rm eff}\ ({\rm K})$ 			& $4893\pm15^{(a)}$ 					& $4822\pm15^{(a)}$ \\
$v_t\ ({\rm km\>s^{-1}})$ 		& $1.48\pm0.05^{(a)}$ 					& $1.27\pm0.05^{(a)}$ \\
$v\sin{i}\ ({\rm km\>s^{-1}})$ 	& $3.36^{(a)}$ 					& $1.55^{(a)}$ \\
${\rm [Fe/H]\ (dex)}$ 			& $-0.77\pm0.03^{(a)}$ 					& $-0.30\pm0.03^{(a)}$ \\
$\log g_{\rm sp}\ ({\rm cgs})$ 		& $2.21\pm0.05^{(a)}$ 			& $2.56\pm0.05^{(a)}$ \\
$M\ (M_{\odot})$ 		& $0.99_{-0.13}^{+0.19}$$^{(b)}$	& $1.47_{-0.20}^{+0.20}$$^{(b)}$ \\
$R\ (R_{\odot})$ 		& $10.64_{-0.59}^{+0.84}$$^{(b)}$ 	& $11.1_{-0.3}^{+1.1}$$^{(b)}$ \\
$\log g_{\rm iso}\ ({\rm cgs})$ 		& $2.42_{-0.10}^{+0.10}$$^{(b)}$ 	& $2.47_{-0.09}^{+0.07}$$^{(b)}$ \\
$\log age\ ({\rm yr})$ 	& $9.84_{-0.22}^{+0.23}$$^{(b)}$ 	& $9.45_{-0.15}^{+0.23}$$^{(b)}$ \\
$\sigma_{\rm HIP}\ ({\rm mag})$		& $0.007$						& $0.008$\\\hline
 \end{tabular}}
\begin{tabnote}
\hangindent6pt\noindent
\hbox to6pt{\footnotemark[$*$]\hss}\unskip%
Each parameter was determined by (a) \citet{Takeda2008} and (b) this work. 
\end{tabnote}
\label{tab1}
\end{table}

\section{OBSERVATION}
All data were obtained with the 1.88-m reflector and High Dispersion Echelle Spectrograph (HIDES; \cite{Izumiura1999}; \cite{Kambe2013}) at Okayama Astrophysical Observatory (OAO).
An iodine absorption cell (I$_{\rm 2}$ cell; \cite{Kambe2002}) was set on the optical path, which superposes numerous iodine absorption lines onto a stellar spectrum in $5000$--$5800\,{\rm \AA}$ as fiducial wavelength reference for precise RV measurements.
In 2007 December, HIDES was upgraded to have three CCDs from one CCD; wavelength coverage of each CCD is $3700$--$5000\,{\rm \AA}$, $5000$--$5800\,{\rm \AA}$, and $5800$--$7500\,{\rm \AA}$.
Therefore, after the upgrade, we evaluate stellar activity level (i.e. Ca\,\emissiontype{II} HK lines) and line profile variations by using the bluest wavelength region simultaneously with stellar RV variations measured by using the middle wavelength region. 

We obtained stellar spectra with two different observation modes of HIDES, the conventional slit mode and the high-efficiency fiber-link mode.
In the slit mode, the slit width was set to $200$ $\micron$ (\timeform{0.76''}), which corresponds to a spectral resolution of 67000 by about 3.3-pixels sampling.
In the fiber-link mode, the width of the sliced image was \timeform{1.05''}, which corresponds to a spectral resolution of 55000 by about 3.8-pixels sampling.
Hereafter, we use slit-spectrum and fiber-spectrum as abbreviation of spectrum taken with slit mode and fiber-link mode, respectively.

The reduction of slit- and fiber-spectrum covering $5000$--$5800\,{\rm \AA}$ and slit-spectrum covering $3700$--$5000\,{\rm \AA}$ was performed with IRAF\footnote{IRAF is distributed by the National Optical Astronomy Observatories, which is operated by the Association of Universities for Research in Astronomy, Inc. under cooperative agreement with the National Science Foundation, USA.} in conventional way.
For fiber-spectrum covering $3700$--$5000\,{\rm \AA}$, the orders in bluer region (shorter than $4000\,{\rm \AA}$) severely overlap with adjacent orders due to using image slicer, and thus we can not subtract scattered light in the region.
Therefore we divided spectrum into red part ($4000$--$5000\,{\rm \AA}$ without order overlapping) and blue part ($3700$--$4000\,{\rm \AA}$ with order overlapping).
In the red part, we can subtract scattered light with IRAF task, apscatter.
In the blue part, we can not subtract scattered light.
Hence we did not perform Ca\,\emissiontype{II} H line analysis for fiber-spectrum.\\
\section{ANALYSIS}
\subsection{Radial velocity}
\label{sec:rv}
For precise RV measurements, we use spectra covering $5000$--$5800\,{\rm \AA}$ on which I$_{\rm 2}$ lines are superimposed by I$_{\rm 2}$ cell.
We compute RV variations, following the method described in \citet{Sato2002}.
We modeled observed spectrum (I$_2$-superposed stellar spectrum) by using stellar template spectrum and high resolution I$_2$ spectrum which are convolved with instrumental profile (IP) of the spectrograph.
The stellar template spectrum was obtained by deconvolving a pure stellar spectrum with the IP estimated from I$_2$-superposed Flat spectrum.
IP was modeled as one central gaussian and ten satellite gaussians where we fix widths and positions of the gaussians and treat their heights as free parameters.
A 2nd-order polynomial is used for the wavelength scaling.
The observed spectrum is divided into hundreds segments with typical width of 150$\>$pixel.
We measured RV values of all the segments and took average of them as the final RV value. 
\subsection{Period search and Keplerian orbital fit}
We first searched for periodic signals in the RV data using generalized Lomb-Scargle periodogram (GLS; \cite{Zechmeister2009}).
GLS can treat each measurement uncertainty as weight and mean RV as a free parameter, which the original Lomb-Scargle periodogram (\cite{Scargle1982}) does not take into account.
The false alarm probability (FAP) of a signal, $p_0$, is defined as
\begin{equation}
FAP=1-\left[1-(1-p_0)^{\frac{N-3}{2}}\right]^M,
\end{equation}
where $N$ is the number of data and $M$ is the number of independent frequencies which we approximated by $M\sim T\Delta f$ ($T$ is a period of observing time and $\Delta f$ is the frequency range we searched; \cite{Cumming2004}).

We derived a Keplerian orbital model for the RV data using the Bayesian Markov chain Monte Carlo (MCMC) method \citep{Gregory2005}.
Fitting parameters are orbital period, $P$, velocity amplitude, $K$, eccentricity, $e$, argument of periastron, $\omega$, orbital phase, $M_0$, RV offset relative to the velocity zero point, $V$, and an additional Gaussian noise including stellar jitter and an unknown noise, $s$.
Since our observations were performed with two distinct modes, RV offset and an additional Gaussian noise are treated as free parameters for each mode.
The prior distribution, minimum and maximum value of each parameter we adopt are shown in Table~\ref{dtab1} (following \cite{Sato2013a}).
We generated $10^7$ points and removed first $10^6$ points to avoid the result affected by an initial condition.
Then we made histogram for each parameter as a PDF, and obtained the median value as the best fit one of the model.
We set the $1\,\sigma$ uncertainty as the range between $15.87$\% and $84.13$\% of the PDF.

\begin{table}
\tbl{Parameter priors for MCMC.}{%
 \begin{tabular}{lcrr}\hline
Parameter & Prior distribution & Minimum & Maximum \\ \hline
$P\ ({\rm d})$ &$[P\ {\rm ln}(P_{\rm max}/P_{\rm min})]^{-1}$ & $1$ & $3000$\\
$K\ (\rm{m\>s^{-1}})$ & $[(K+K_a)\ {\rm ln}[(K_a+K_{\rm max})/K_a]]^{-1}$ & $0\ (K_a=5)$ & $500$\\
$e$ & $[e_{\rm max}-e_{\rm min}]^{-1}$ & $0$ & $1$\\
$\omega$ & $[\omega_{\rm max}-\omega_{\rm min}]^{-1}$ & $0$ & $2\pi$\\
$M_0$ & $[M_{\rm max}-M_{\rm min}]^{-1}$ & $0$ & $2\pi$\\
$V\ (\rm{m\>s^{-1}})$ & $[V_{\rm max}-V_{\rm min}]^{-1}$ & $-100$ & $100$\\
$s\ (\rm{m\>s^{-1}})$ & $[(s+s_a)\ {\rm ln}[(s_a+s_{\rm max})/s_a]]^{-1}$ & $0\ (s_a=1)$ & $100$\\ \hline
 \end{tabular}}
\label{dtab1}
\end{table}

\subsection{Chromospheric activity}
RV variations could also be explained by stellar chromospheric activity.
Therefore, we checked flux variability of the core of chromospheric Ca\,\emissiontype{II} H line.
Ca\,\emissiontype{II} K line was not used for the analysis because of the low $S/N$ in bluer wavelength region.
Following \citet{Sato2013a}, we define Ca\,\emissiontype{II} H index, $S_{\rm{H}}$, as
\begin{equation}
S_{\rm{H}}=\frac{F_{\rm{H}}}{F_{\rm{B}}+F_{\rm{R}}},
\end{equation}
where $F_{\rm H}$ is the integrated flux within $0.66\,{\rm \AA}$ wide bin centered at Ca\,\emissiontype{II} H line, and $F_{\rm B}$, $F_{\rm R}$ is $1.1\,{\rm \AA}$ wide bin centered at $\pm1.2\,{\rm \AA}$ from Ca\,\emissiontype{II} H line core, respectively.
The error was estimated based on photon noise.
If stellar chromospheric activity causes an apparent RV variation, we can see a correlation between RV and $S_{\rm{H}}$ (e.g., Figure~9 in \cite{Sato2013a}).
Thus we can evaluate the effect of the stellar activity on RV by $S_{\rm{H}}$.
The spectrum near Ca\,\emissiontype{II} H line of 24 Boo and $\gamma$ Lib are shown in Figure~\ref{re_cahk_127_138}, from which we see no significant reversal emission in the line core.
\subsection{Line profile analysis}
\label{sec:BIS}
The spectral line-profile deformation causes the apparent wavelength shift which may masquerade as a planetary signal.
To evaluate this possibility, we performed the line profile analysis.
Although it is ideal that the same wavelength regions are used for both of the line profile analysis and the RV measurements, the wavelength range of 5000--5800$\,{\rm \AA}$ used for RV analysis is contaminated by I$_{\rm 2}$ lines and thus unsuitable for the line profile analysis. 
Therefore we used iodine-free spectra in the range of $4000$--$5000\,{\rm \AA}$ for the line profile analysis.

Our analysis method is based on that adopted in Queloz et al (2001). 
First we calculate the weighted cross-correlation function (CCF; \cite{Pepe2002}) of a stellar spectrum and a numerical mask, which corresponds to a mean profile of stellar absorption lines in the relevant wavelength range. 
The numerical mask was based on a model spectrum created with use of SPECTRUM\footnote{\tt{http://www.appstate.edu/\%7egrayro/spectrum/spectrum.html}} \citep{Gray1994} for a G-type giant. 
Next we calculate the bisector inverse span (BIS; \cite{Dall2006}) of CCF as
\begin{equation}
{\rm BIS} = v_{top} - v_{bot},
\end{equation}
where $v_{top}$ is the averaged velocity of bisector's upper region ($5$\%--$15$\% from the continuum of CCF) and $v_{bot}$ is the averaged velocity of bisector's lower region ($85$\%--$95$\% from the continuum of CCF).
Thus defined BIS can be a measure of line-profile asymmetry.
\section{RESULTS}
\subsection{24 Boo (HR 5420, HD 127243, HIP 70791)}
\begin{figure}
\begin{center}
\includegraphics[width=8cm]{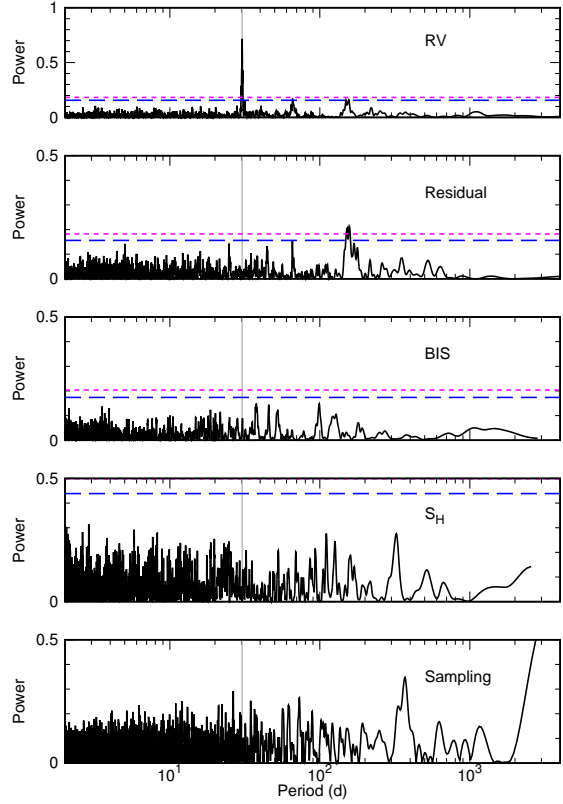}
\end{center}
\caption{The GLS periodograms for 24 Boo. From top to bottom: the observed RVs, the RV residuals to the single Keplerian fitting, BIS, $S_{\rm{H}}$ and data sampling, respectively. The blue and magenta dotted lines show 1\% and 0.1\% FAP of the periodogram, respectively. The gray vertical line shows the $30\>$d periodicity detected in the RV data.}
\label{peri_127243}
\end{figure}

\begin{figure}
\begin{center}
\includegraphics[width=8cm]{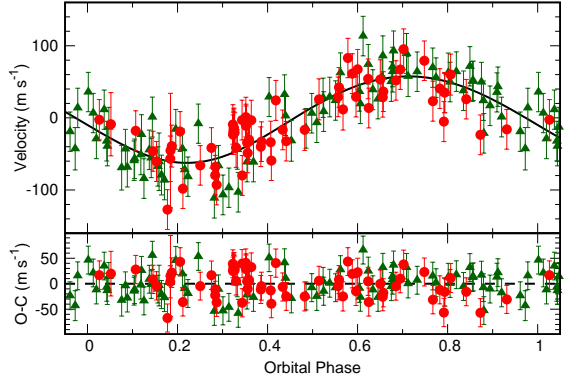}
\end{center}
\caption{Top: The observed RVs and the best-fit single Keplerian curve for 24 Boo. The horizontal axis is the orbital phase. The derived stellar jitter is also included in the error bar. Red circles and green triangles are data taken with slit- and fiber-mode, respectively. Bottom: The RV residuals to the orbital fitting.}
\label{re_rv_resi_127243}
\end{figure}

We collected a total of 149 RV data for 24 Boo between 2003 April and 2016 June, and they are listed in Table~\ref{rvtab_127}.
Firstly we fitted a single Keplerian curve to the data with a period of $30\>$d, which has the strongest peak in the periodogram (Figure~\ref{peri_127243}).
By performing the MCMC fitting, we obtained orbital parameters of $P= 30.3506_{-0.0077}^{+0.0078}\>$d, $K=59.9_{-3.2}^{+3.3}\>{\rm m\>s^{-1}}$ and $e=0.042_{-0.029}^{+0.048} $.
Adopting a stellar mass of $0.99\,M_{\odot}$, we obtained a minimum mass $m_p\sin{i}=0.91_{-0.10}^{+0.13}\,M_{\rm Jup}$ and a semimajor axis $a=0.19_{-0.009}^{+0.012}\>{\rm au}$ for the planet.
The orbital parameters we derived are listed in Table~\ref{dtab2}.
Due to the small eccentricity, the longitude of periastron and the time of periastron have large uncertainties.
The resulting single Keplerian curve and the observed RVs are shown in Figure~\ref{re_rv_resi_127243}.

\begin{figure}
\begin{center}
\includegraphics[width=10cm]{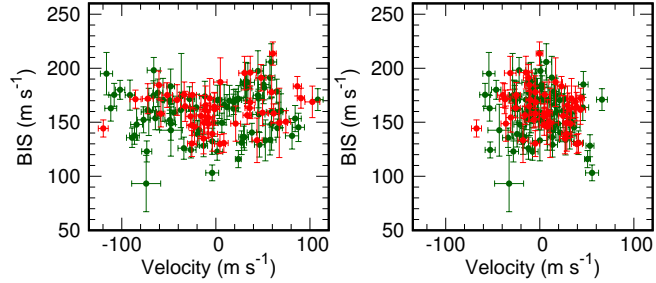}
\end{center}
\caption{Left: BIS plotted as a function of the observed RV for 24 Boo. Red and green circles are data taken with slit- and fiber-mode, respectively. Right: BIS plotted as a function of the RV residuals to the single Keplerian fitting.}
\label{re_rv_resi_bis_127243}
\end{figure}

\begin{figure}
\begin{center}
\includegraphics[width=10cm]{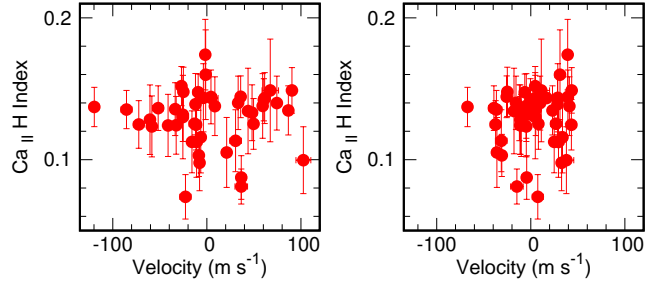}
\end{center}
\caption{Left: $S_{\rm H}$ plotted as a function of the observed RV for 24 Boo. $S_{H}$ was obtained only from slit-spectrum. Right: $S_{\rm H}$ plotted as a function of the RV residuals to the single Keplerian fitting.}
\label{re_rv_bis_cahk_127243}
\end{figure}

To investigate the possible cases which cause apparent RV variations, we evaluate BIS and $S_{\rm H}$.
We calculated BIS for the data taken with both slit- and fiber-mode (Figure~\ref{re_rv_resi_bis_127243}).
We found that line-profile of slit-spectra tended to vary greatly than that of fiber-spectra especially for slowly rotating stars ($v\sin{i}\lesssim3\>{\rm km\>s^{-1}}$) due to IP variations (see Appendix).
In the case of 24 Boo, which has a $v\sin{i}=3.36\>{\rm km\>s^{-1}}$, BIS value of slit-spectra and fiber-spectra show almost the same level of dispersion (Figure~\ref{re_rv_resi_bis_127243} and Figure~\ref{rv_bis_ip_127_138}; RMSs of BIS are $18.5\>{\rm m\>s^{-1}}$ and $22.3\>{\rm m\>s^{-1}}$ for slit-spectra and fiber-spectra, respectively).
Therefore stellar surface modulation has a large effect on the spectral line-profile of the star.
Nevertheless, BIS values have no correlation with RV (correlation coefficient, $r=0.18$ and $-0.09$ for slit- and fiber-spectra, respectively; Figure~\ref{re_rv_resi_bis_127243}).
We calculated $S_{\rm H}$ for the data taken with only slit-mode, and the $S_{\rm H}$ values also have no correlation with RV ($r=-0.09$; Figure~\ref{re_rv_bis_cahk_127243}).
Furthermore BIS and $S_{\rm H}$ show no significant periodicity in the periodograms (Figure~\ref{peri_127243}).
Thus we can conclude the RV variations with a period of 30$\>$d should be caused by orbital motion.

As seen in Figure~\ref{peri_127243}, the RV residuals to the single Keplerian fitting exhibited possible periodicity of $\sim150\>$d.
Thus we applied double Keplerian fitting to the RV data.
As a result, RMS of the residuals was improved by $3\>{\rm m\>s^{-1}}$, and we obtained orbital parameters of $P= 152.1_{-0.9}^{+5.0}\>$d, $K = 18.7_{-3.5}^{+3.7}\>{\rm m\>s^{-1}}$, $e=0.28_{-0.18}^{+0.20}$ and $m_p\sin{i}=0.504_{-0.092}^{+0.094}\,M_{\rm Jup}$ for the possible second planet.
To evaluate the validity of introducing the second planet, we calculated BIC (Bayesian information criterion) value for both of the single- and the double-planet cases as, ${\rm BIC}=\chi^2+n\ln{(N_{\rm data})}$ where $n$ is the number of free parameters and $N_{\rm data}$ is the number of data points.
When we fixed stellar jitter to $25\>{\rm m\>s^{-1}}$ which is estimated based on the residuals, BIC were 184 for the double-planet model and 196 for the single-planet model.
Although the second planet is statistically possible, $\Delta$BIC becomes less significant when we assume larger stellar jitter.
Combined with the fact that the maximum rotational period of 24 Boo is about $160\>$d, which is close to the detected periodicity in the RV residuals, we can not certainly confirm the existence of the second planet at this stage.
The expected RV amplitude of stellar oscillations, $v_{\rm osc}$, derived with the scaling relations \citep{Kjeldsen1995} for the star is $14\>{\rm m\>s^{-1}}$.
Given the expected $v_{\rm osc}$ and widely varying BIS, stellar oscillations combined with stellar activity should be plausible explanation for the large RMS of the residuals to the single Keplerian fitting.


\begin{table}
\tbl{Radial velocities of 24 Boo.\footnotemark[$*$]}{%
 \begin{tabular}{cccc}\hline
JD 				& Radial Velocity 		& Uncertainty 		& Observation mode\\ 
($-2450000$) 				& $(\rm{m\>s^{-1}})$		& $(\rm{m\>s^{-1}})$ 		& \\ \hline
$2739.2188$ & $-9.2$ & $8.8$ & slit \\
$3052.1609$ & $42.5$ & $6.3$ & slit \\
$3162.1789$ & $31.6$ & $6.0$ & slit \\
$3366.3486$ & $-38.8$ & $6.0$ & slit \\
$3403.3274$ & $4.1$ & $6.0$ & slit \\ \hline
 \end{tabular}}
 \begin{tabnote}
\hangindent6pt\noindent
\hbox to6pt{\footnotemark[$*$]\hss}\unskip%
Only first 5 data are listed. All data are presented in the electronic table. 
\end{tabnote}
\label{rvtab_127}
\end{table}

\begin{table}
\tbl{Orbital parameters.}{%
 \begin{tabular}{lccccc}\hline
Parameter 						& 24 Boo\ b 				& $\gamma$ Lib\ b 				& $\gamma$ Lib\ c 	& $\gamma$ Lib\ b 				& $\gamma$ Lib\ c\\ 
								&Keplerian model		 &\multicolumn{2}{c}{Keplerian model}		& \multicolumn{2}{c}{Dynamical fitting}\\ \hline
$P\ (\rm{d})$ 						&$ 30.3506_{-0.0077}^{+0.0078} $ 	&$ 415.2_{-1.9}^{+1.8} $ 			&$ 964.6_{-3.1}^{+3.1} $		&--						&--\\
$K\ (\rm{m\>s^{-1}})$ 					&$ 59.9_{-3.2}^{+3.3} $			&$ 22.0_{-2.2}^{+2.2} $			&$ 73.0_{-2.1}^{+2.1} $		&--						&--\\
$e$ 								&$ 0.042_{-0.029}^{+0.048} $		&$ 0.21_{-0.10}^{+0.10} $ 		&$ 0.057_{-0.032}^{+0.034} $	&$ 0.25^{+0.020}_{-0.031} $	&$0.054_{-0.022}^{+0.022}$\\
$\omega\ (^{\circ})$ 					&$ 210_{-130}^{+100} $			&$ 187_{-38}^{+30} $			&$ 146_{-43}^{+32} $		&$219.1 ^{+6.6}_{-10.9}$		&$154.3^{+9.3}_{-7.2}$\\
$T_p\ (\rm{JD-2450000})$ 			&$ 8.6_{-7.9}^{+10.1}$ 		&$ 41_{-33}^{+36} $			&$ 725_{-110}^{+87} $		&$47^{+14}_{-16}$			&$716^{+11}_{-16}$\\
$m_p\sin{i}\ (M_{\rm Jup})$ 		&$ 0.91_{-0.10}^{+0.13} $ 		&$ 1.02_{-0.14}^{+0.14} $			&$ 4.58_{-0.43}^{+0.45} $		&$1.029^{+0.087}_{-0.087}$	&$4.63^{+0.22}_{-0.12}$\\
$a\ ({\rm au})$ 						&$ 0.190_{-0.009}^{+0.012} $ 	&$ 1.24_{-0.10}^{+0.10} $ 		&$ 2.17_{-0.10}^{+0.10} $			&$1.23968^{+0.00066}_{-0.00070}$	&$2.1761^{+0.0025}_{-0.0020}$\\ \hline
$s_{\rm{slit}}\ (\rm{m\>s^{-1}})$ 			&$ 26.4_{-2.5}^{+2.9} $ 			& \multicolumn{2}{c}{$ 17.6_{-1.3}^{+1.5} $} 					&\multicolumn{2}{c}{$17.6$ (fixed)}\\ 
$s_{\rm{fiber}}\ (\rm{m\>s^{-1}})$ 		&$ 27.8_{-2.8}^{+3.3} $ 			& \multicolumn{2}{c}{$ 13.8_{-1.7}^{+2.1} $} 					&\multicolumn{2}{c}{$13.8$ (fixed)}\\ 
$v_{\rm{offset, slit}}\ (\rm{m\>s^{-1}})$ 	&$ 8.1_{-3.9}^{+4.0} $ 			& \multicolumn{2}{c}{$ 6.5_{-1.8}^{+1.9} $} 					&\multicolumn{2}{c}{$6.5$ (fixed)}\\ 
$v_{\rm{offset, fiber}}\ (\rm{m\>s^{-1}})$ 	&$ 7.0_{-4.1}^{+4.1} $ 			& \multicolumn{2}{c}{$ 25.4_{-2.7}^{+2.7} $} 					&\multicolumn{2}{c}{$25.4$ (fixed)}\\ \hline
$\rm{RMS}\ (\rm{m\>s^{-1}})$ 			& 26.51 & \multicolumn{2}{c}{16.41} 														&\multicolumn{2}{c}{16.40}\\
${N}_{\rm obs, slit}$ 			& 67 							& \multicolumn{2}{c}{106} & \multicolumn{2}{c}{106}\\
${N}_{\rm obs, fiber}$ 		& 82 							& \multicolumn{2}{c}{40} & \multicolumn{2}{c}{40}\\\hline
 \end{tabular}}
 \begin{tabnote}
\hangindent6pt\noindent
 \hbox to6pt{\footnotemark[$*$]\hss}\unskip%
The uncertainties of host stars mass are included in those of $m_p\sin{i}$ and $a$ for Keplerian orbital parameters. 
To obtain dynamical orbital parameters of $\gamma$ Lib b and c, we adopt a stellar mass of $1.47\,M_{\odot}$ without including the uncertainty.
The dynamical orbital parameters are those at JD$=$2450000.
\end{tabnote}
\label{dtab2}
\end{table}

\subsection{$\gamma$ Lib (HR 5787, HD 138905, HIP 76333)}\label{res_gamma}
\begin{figure}
\begin{center}
\includegraphics[width=8cm]{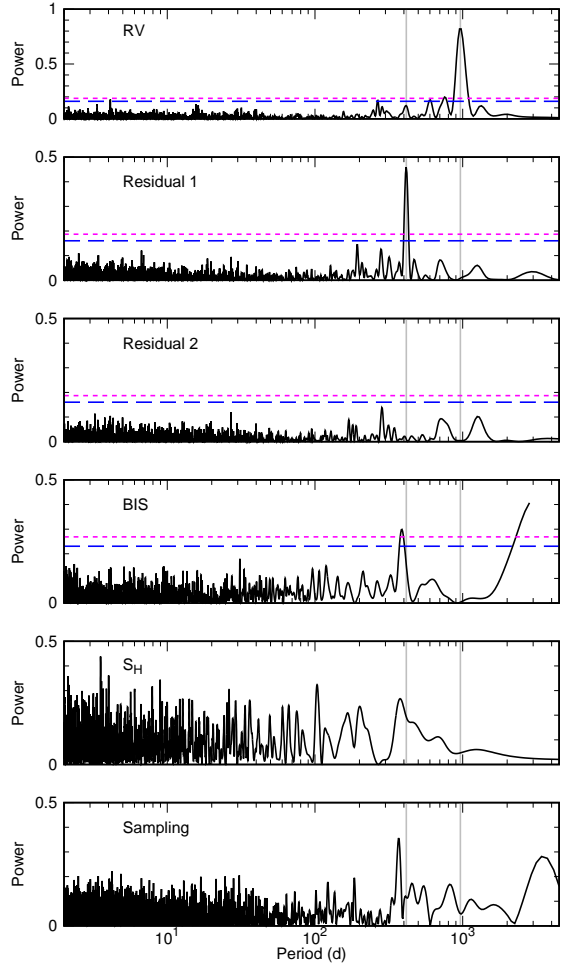}
\end{center}
\caption{The GLS periodograms for $\gamma$ Lib. From top to bottom: the observed RVs, the RVs without that of $\gamma$ Lib c (Residual 1), the RV residuals to the double Keplerian fitting (Residual 2), BIS, $S_{\rm H}$ and data sampling, respectively. The blue and magenta dotted lines show $1$\% and $0.1$\% FAP of the periodogram, respectively. The gray vertical lines show $415\>$d and $964\>$d periodicities.}
\label{peri_138905}
\end{figure}

\begin{figure}
\begin{center}
\includegraphics[width=8cm]{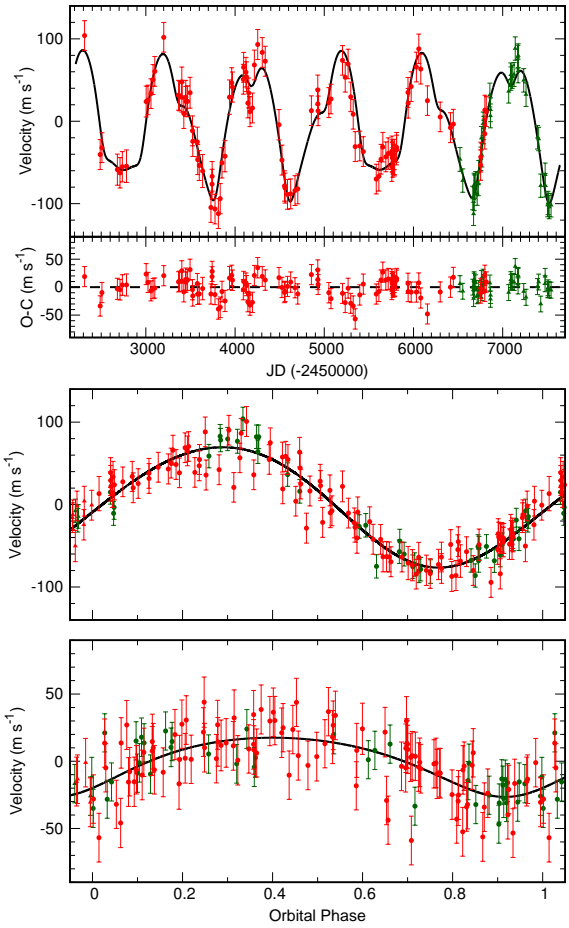}
\end{center}
\caption{Top: The observed RVs and the best-fit double Keplerian curve for $\gamma$ Lib. The horizontal axis is JD. The derived stellar jitter is also included in the error bar. Red circles and green triangles are data taken with slit- and fiber-mode, respectively. The RV residuals to the double Keplerian fitting are plotted in the lower part of the panel. Middle: The phase-folded RV and single Keplerian curve for $\gamma$ Lib c. Bottom: The phase-folded RV and single Keplerian curve for $\gamma$ Lib b.}
\label{re_rv_resi_138905}
\end{figure}

We collected a total of 146 RV data for $\gamma$ Lib between 2002 February and 2016 June, and they are listed in Table~\ref{rvtab_138}.
Firstly we fitted a single Keplerian curve to the data with a period of $964\>$d, at which the strongest peak is confirmed in the periodogram (Figure~\ref{peri_138905}).
As seen in Figure~\ref{peri_138905}, the residuals to the single Keplerian fitting exhibited significant periodicity of $\sim415\>$d.
Thus we applied double Keplerian fitting to the RV data, and RMS of the RV residuals was improved by $6\>{\rm m\>s^{-1}}$.
In the same way as section 5.1, we calculated BIC values for both of the single- and the double-planet cases.
When we fixed stellar jitter to $20\>{\rm m\>s^{-1}}$ which is estimated based on the residuals, BIC were 154 for double-planet case and 218 for single-planet case.
The resulting $\Delta$BIC becomes 64, and such a large value indicates that the possible second planet is statistically favored.
By performing the MCMC fitting, we obtained orbital parameters, and they are listed in Table~\ref{dtab2}.
The orbital parameters of the inner planet ($\gamma$ Lib b) are $P= 415.2_{-1.9}^{+1.8}\>$d, $K = 22.0_{-2.2}^{+2.2}\>{\rm m\>s^{-1}}$ and $e=0.21_{-0.10}^{+0.10}$, and those of the outer planet ($\gamma$ Lib c) are $P= 964.6_{-3.1}^{+3.1}\>$d, $K=73.0_{-2.1}^{+2.1}\>{\rm m\>s^{-1}}$ and $e=0.057_{-0.032}^{+0.034}$.
Adopting a stellar mass of $1.47\,M_{\odot}$, we obtained a minimum mass of $m_p\sin{i}=1.02_{-0.14}^{+0.14}$, $4.58_{-0.43}^{+0.45}\,M_{\rm Jup}$ and a semimajor axis $a=1.24_{-0.10}^{+0.10}$, $2.17_{-0.10}^{+0.10}\>{\rm au}$ for inner and outer planet, respectively.
The orbital parameters we obtained are listed in Table~\ref{dtab2}.
The resulting double Keplerian curve and the observed RVs are shown in Figure~\ref{re_rv_resi_138905}.

\begin{figure}
\begin{center}
\includegraphics[width=14cm]{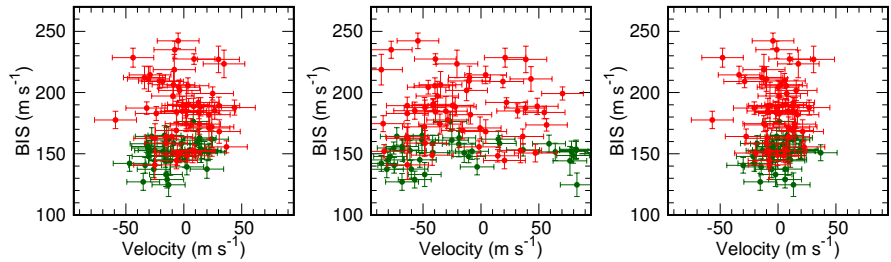}
\end{center}
\caption{Left: BIS plotted as a function of the RV of the inner planet for $\gamma$ Lib. Red and green circles are data obtained from slit and fiber-spectrum, respectively. Middle: BIS plotted as a function of the RV of the outer planet for $\gamma$ Lib. Right: BIS plotted as a function of the RV residuals to the double Keplerian fitting.}
\label{re_rv_resi_bis_138905}
\end{figure}

\begin{figure}
\begin{center}
\includegraphics[width=14cm]{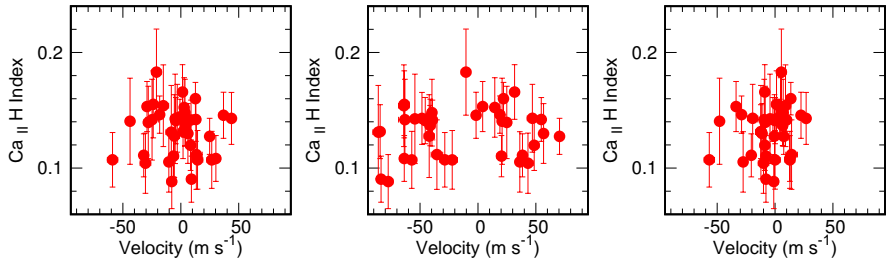}
\end{center}
\caption{Left: $S_{\rm H}$ plotted as a function of the RV of the inner planet for $\gamma$ Lib. $S_{\rm H}$ was obtained only from slit-spectrum. Middle: $S_{\rm{H}}$ plotted as a function of the RV of the outer planet for $\gamma$ Lib. Right: $S_{\rm H}$ plotted as a function of the RV residuals to the double Keplerian fitting.}
\label{re_rv_cahk_138905}
\end{figure}

\begin{figure}
\begin{center}
\includegraphics[width=8cm]{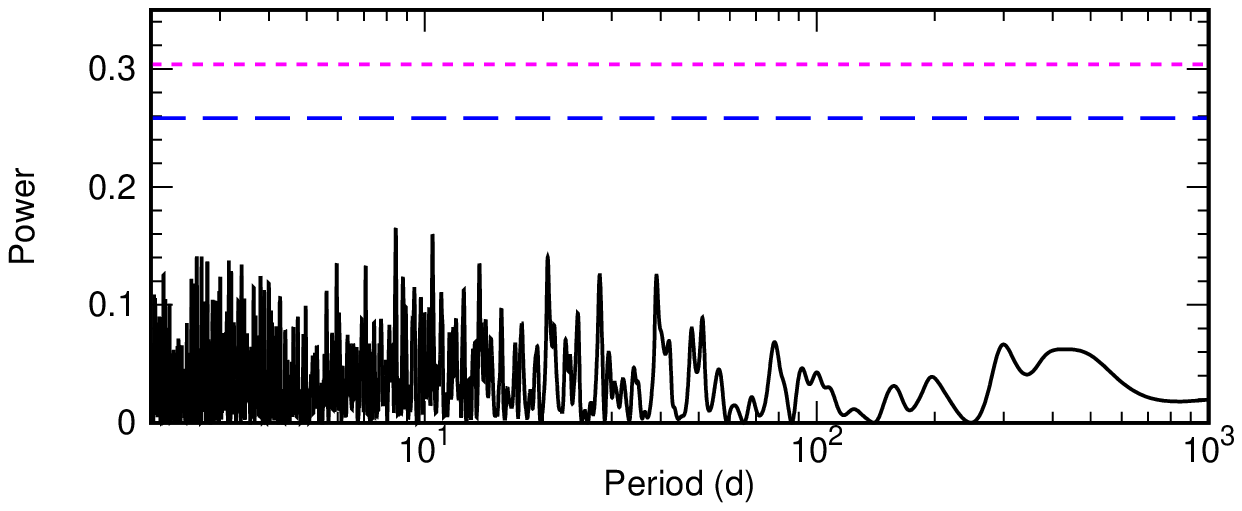}
\end{center}
\caption{GLS periodogram of $\gamma$ Lib Hipparcos photometry. The blue and magenta dotted lines shows $1$\% and $0.1$\% FAP of the periodogram, respectively.}
\label{re_peri_hip_138905}
\end{figure}

The small velocity amplitude of $22\>{\rm m\>s^{-1}}$ for the inner planet could easily be caused by stellar activity, and thus we investigate the possible case which causes apparent RV variations.
In the case of this star, we found that BIS of the slit-spectra varied greatly than that of fiber-spectra (Figure~\ref{re_rv_resi_bis_138905} and Figure~\ref{rv_bis_ip_127_138}).
This means that BIS variations for $\gamma$ Lib are dominated by the IP variability.
Although BIS values have a significant periodicity at $385\>$d in the periodogram, this is possibly a 1-year or sampling alias.
$S_{\rm H}$ values have no correlation with RV ($r=-0.08$ and $0.27$ for slit- and fiber-spectra, respectively; Figure~\ref{re_rv_cahk_138905}), and they show no significant periodicity in the periodogram (Figure~\ref{peri_138905}).
We also calculated periodogram of Hipparcos photometry, and found no significant peaks in the periodogram (Figure~\ref{re_peri_hip_138905}).
Combined with these facts, we conclude that the RV variations with period of $415\>$d as well as $964\>$d are best explained by orbital motion.
Although the RMS of the residuals to the double Keplerian fitting ($16\>{\rm m\>s^{-1}}$) is still large compared to typical RV uncertainties ($\sim4.4\>{\rm m\>s^{-1}}$), 
the value is comparable to the expected RV amplitude of stellar oscillation for the star ($11.3\>{\rm m\>s^{-1}}$; \cite{Kjeldsen1995}).

\begin{table}
\tbl{Radial velocities of $\gamma$ Lib.\footnotemark[$*$]}{%
 \begin{tabular}{cccc}\hline
JD 				& Radial Velocity 		& Uncertainty 		& Observation mode\\ 
($-2450000$) 				& $(\rm{m\>s^{-1}})$		& $(\rm{m\>s^{-1}})$ 		& \\ \hline
$2313.3525$ & $110.5$& $4.0$ & slit \\
$2488.0122$ & $-33.9$ & $4.6$ & slit \\
$2507.9571$ & $-25.5$ & $4.3$ & slit \\
$2680.3842$ & $-52.2$ & $6.9$ & slit \\
$2710.3387$ & $-56.1$ & $5.5$ & slit \\ \hline
 \end{tabular}}
 \begin{tabnote}
\hangindent6pt\noindent
\hbox to6pt{\footnotemark[$*$]\hss}\unskip%
Only first 5 data are listed. All data are presented in the electronic table. 
\end{tabnote}
\label{rvtab_138}
\end{table}

\section{DISCUSSION}
We here reported two new planetary systems around evolved stars: 24 Boo and $\gamma$ Lib.
This result is based on RV measurements performed at OAO with the 1.88-m reflector and HIDES.
We evaluate stellar activity level (Ca\,\emissiontype{II} H line) and line profile variations, and we conclude that the periodic RV variations are caused by planetary orbital motion.
The solar-mass giant star 24 Boo ($0.99\,M_{\Sol}$, $10.64\,R_{\Sol}$, and $\log g=2.42$) has a planet with a semimajor axis of $a=0.19_{-0.009}^{+0.012}\>{\rm au}$.
The intermediate-mass giant star $\gamma$ Lib ($1.47\,M_{\Sol}$, $11.1\,R_{\Sol}$, and $\log g=2.47$) has two planets with semimajor axes of $a=1.24_{-0.10}^{+0.10}$ and $2.17_{-0.10}^{+0.10}\>{\rm au}$, respectively.
We note that both stars have low metallicities (${\rm [Fe/H]}=-0.77$ for 24 Boo and $-0.30$ for $\gamma$ Lib) among the giant stars hosting planetary companions.

\subsection{Planetary semimajor axis}
\label{sub:Planetsemi}
\begin{figure}
\begin{center}
\includegraphics[width=8cm]{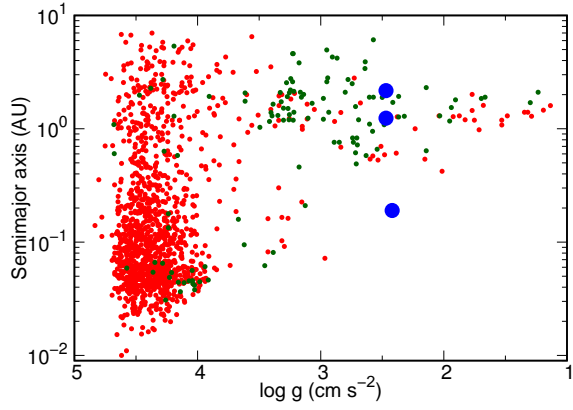}
\end{center}
\caption{
Semimajor axis of planets as a function of $\log g$ of planet host stars. Data are taken from Exoplanets.eu (http://exoplanet.eu).
Red filed circles and green filled circles represent solar-mass stars with $0.7$--$1.5\,M_{\Sol}$ and intermediate-mass stars with $>1.5\,M_{\Sol}$, respectively.
24 Boo b and $\gamma$ Lib b and c are indicated with blue circles.}
\label{re_logg_a}
\end{figure}

\begin{figure}
\begin{center}
\includegraphics[width=8cm]{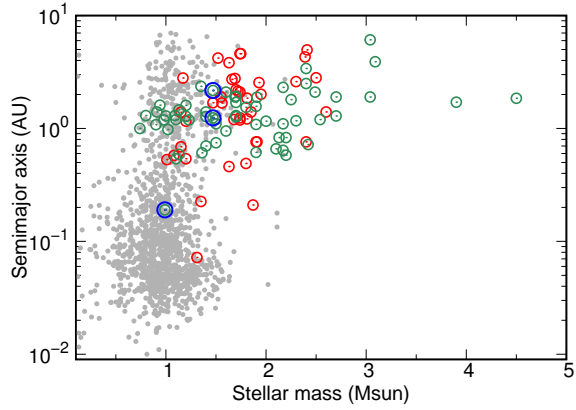}
\end{center}
\caption{Semimajor axis of planets as a function of stellar mass. Data are taken from Exoplanets.eu (http://exoplanet.eu). Green open circle, red open circle and gray filled circle represent stars with stellar radii of $R>10\,R_{\solar}$, those with stellar radii of $5\,R_{\solar}<R<10\,R_{\solar}$ and those with stellar radii of $R< 5\,R_{\solar}$, respectively. 24 Boo b and $\gamma$ Lib b and c are indicated with blue circles.}
\label{mass_semi}
\end{figure}

24 Boo b has the shortest orbital period ever found around evolved stars with stellar radius larger than $10\,R_{\solar}$. 
On the other hand, $\gamma$ Lib b and c have rather typical orbital periods which follow other planets around giant stars, whereas their orbital configuration is worth discussing (see section 6.3).
TYC-3667-1280-1 b ($P=26.4\>{\rm d},\ a=0.21\>{\rm au}$; \cite{Niedzielski2016}) and Kepler-91 b ($P=6.2\>{\rm d},\ a=0.072\>{\rm au}$; \cite{Lillo-Box2014a}, \yearcite{Lillo-Box2014b}; \cite{Barclay2015}; \cite{Sato2015}) are two of the shortest-period planets ever found around giant stars, whose stellar radii are about $5-6\,R_{\solar}$.
Figure~\ref{re_logg_a} shows a distribution of the semi-major axis of currently known planets plotted against their host star's surface gravity\footnote{
For Figure~\ref{re_logg_a}, \ref{mass_semi} and \ref{feh_semi}, data are taken from Exoplanets.eu (http://exoplanet.eu) except for Kepler-433 and BD+20 2457. 
We refer stellar parameters of Kepler-433 from \citet{Almenara2015}, since incorrect parameters are listed. 
We refer stellar parameters of BD+20 2457 from \citet{Mortier2013}. 
However BD+20 2457 still have the lowest metallicity among the giant stars, and this would not change the content of following discussion.}. 
As seen in the figure, there are few planets which reside at $a<0.5\>{\rm au}$ around giant stars.
We also show a distribution of the semi-major axis of currently known planets against their host star's mass in Figure~\ref{mass_semi}. 
In the figure, we can see that solar-mass giant stars with stellar radii of $R>5\,R_{\solar}$ have few short-period planets, which are commonly found around solar-mass stars at the main-sequence stage.
Additionally, we can see that less evolved massive stars, which are progenitors of massive giant stars, have short-period planets.
Thus the stellar evolution should be a key factor for the lack of short-period planets around giant stars.
It should be noted that there might be an observational bias in the planetary orbital distribution around massive stars.
While giant planets around massive giant stars which reside at $a>0.2\>{\rm au}$ were mainly detected by RV method, most of the short-period planets around less evolved massive stars are detected by transit method.
Furthermore \citet{Jones2017arXiv} showed that masses of short-period planets around giant stars ($\log g\lesssim3.5$) are largely close to or lower than $1.0\,M_{\rm Jup}$, and such low mass planets should be difficult to be detected by RV method.
24 Boo is highly evolved ($\log g=2.42$) compared to other evolved stars with short-period planets, and the companion is expected to have experienced tidal orbital decay ($a/R_\star=3.84$) \citep{Villaver2014}.
The influence of stellar evolution has been intensively investigated for stars with masses larger than $1.5\,M_{\solar}$, and thus 24 Boo and its companion have an important role in understanding the influence of stellar evolution for low-mass stars.

As shown in Figure~\ref{re_HR}, we can see that 24 Boo is expected to be ascending the red giant branch (RGB).
A stellar radius of one solar-mass star is expected to reach $1\>{\rm au}$ at the tip of the RGB (e.g., \cite{Villaver2009}).
Therefore 24 Boo b will be engulfed by the central star before it reaches the tip of the RGB.
In contrast, since the planets orbiting $\gamma$ Lib are far apart from the central star, $\gamma$ Lib b and c will survive engulfment even if the star reached the tip of the RGB.

Given the short orbital period of 24 Boo b and host star's large radius of $10.64\,R_{\solar}$, a probability of transit for the planet is $25$\% and the transit duration is about $60\>$h.
The incident flux of $\sim4.3\times10^8\>{\rm erg\>s^{-1}\>cm^{-2}}$, which corresponds to an equilibrium temperature of $1770^{+90}_{-70}\>{\rm K}$ assuming a Bond albedo of 0.0, exceeds a planet inflation threshold of $2\times10^8\>{\rm erg\>s^{-1}\>cm^{-2}}$ suggested by \citet{Demory2011}, and thus the planet is expected to be inflated (\cite{Lopez2016}). 
Assuming the planetary radius to be between $1$ and $2\,R_{\rm Jup}$, which corresponds to the typical radius of well-characterized hot Jupiters ever found \citep{Grunblatt2016}, a transit depth is expected to be $0.01$--$0.04$\% and it is detectable by space telescopes.
\subsection{Planet-metallicity relation}
\begin{figure}
\begin{center}
\includegraphics[width=8cm]{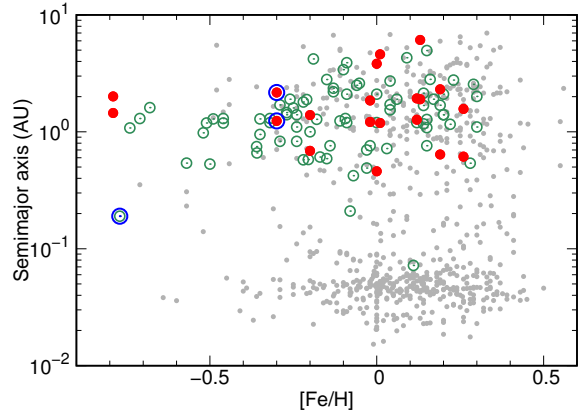}
\end{center}
\caption{
Semimajor axis of planets with masses larger than $0.1\,M_{\rm Jup}$ as a function of stellar metallicity ([Fe/H]). 
Data are taken from Exoplanets.eu (http://exoplanet.eu). 
Green open circles and red filled circles represent giant stars ($R>5\,R_{\solar}$) with single-planet and multiple-planet, respectively, and gray filled circles represent others. 
24 Boo b and $\gamma$ Lib b and c are indicated with blue circles}
\label{feh_semi}
\end{figure}

Figure~\ref{feh_semi} shows the semi-major axis of currently known planets with masses larger than $0.1\,M_{\rm Jup}$ against their host star's metallicity.
From this figure, we can see that 24 Boo and $\gamma$ Lib are relatively metal-poor stars.
24 Boo has the second lowest metallicity among the giant stars with stellar radii larger than $5\,R_{\solar}$ after BD+20 2457 ($M=1.06\,M_{\Sol}$, $R=33.0\,R_{\Sol}$, and ${\rm [Fe/H]}=-0.79$; \cite{Mortier2013}), which hosts a double brown-dwarf system.
Furthermore, compared to the planets around the stars with ${\rm [Fe/H]}<-0.70$, 24 Boo b has the shortest orbital period.
$\gamma$ Lib has the second lowest metallicity among the giant stars with multiple-planetary systems after BD+20 2457 (\cite{Niedzielski2009}).

From the following viewpoints, short-period giant planets orbiting metal-poor stars are expected to be rare.
Firstly, giant planets are generally hard to be formed around metal-poor stars.
The RV surveys focusing on evolved stars (\cite{Reffert2015}, \cite{Jones2016} and \cite{Wittenmyer2017}) showed a positive correlation between stellar metallicity and occurrence rate of giant planets, 
which has been already observed for the cases of main-sequence stars (\cite{Fischer2005} and \cite{Udry2007}).
This is consistent with one of the well-established planet formation scenario, core-accretion model (e.g., \cite{Pollack1996}).
The core-accretion model requires solid materials in the proto-planetary disk to form planets, and thus the stellar metallicity relates with a planet formation rate.
Secondly, the stellar metallicity might affect a planetary orbital distance.
Since the opacity of metal-poor proto-planetary disk is thin, the disk is expected to diffuse rapidly by photoevaporation (\cite{Ercolano2010}).
This results in a short disk lifetime, and planet inward migration would be halted.
Indeed, \citet{Dawson2013} investigated planet-metallicity relation and showed that, among planets detected by Kepler space telescope, metal-rich (${\rm [Fe/H]}>0.0$) stars tended to host more short-period planets ($a<0.1\>{\rm au}$) than metal-poor stars (${\rm [Fe/H]}<0.0$).
We note that the stellar samples used in \citet{Dawson2013} largely account for main-sequence stars.
\citet{Jofre2015} compiled the evolved stars with planets detected via RV method and claimed a similar relation for the case of subgiant stars\footnote{In \citet{Jofre2015}, stars with $M_{\rm bol}<2.82$ are classified as giant stars and those with $M_{\rm bol}>2.82$ are classified as subgiant stars.
They showed that planets orbiting at $a<0.5\>{\rm au}$ were found around subgiant stars with ${\rm [Fe/H]}>0.0$, while planets orbiting at $a>0.5\>{\rm au}$ were found around subgiant stars with various metallicities including subsolar values.
On the other hand, planets orbiting at $a<0.5\>{\rm au}$ are rarely found around giant stars independently of stellar metallicity.
Although this could come from stellar mass and/or evolution as well as metallicity, we have no clear answer at this stage due to the small number of confirmed planets around giant stars.
24 Boo b would help to make clear the process of planetary formation and migration around metal-poor stars.}

A relation between stellar metallicity and multiplicity of planets have not been clarified especially for evolved stars.
\citet{Jofre2015} showed that multiple-planet systems of evolved stars were more metal-rich than single-planet systems of evolved stars by $\sim0.13\>{\rm dex}$, which followed the case of main-sequence stars \citep{Wright2009}.
Though we can see certainly same trend in Figure~\ref{feh_semi}, the number of sample is not enough to evaluate the relation.
Since $\gamma$ Lib has very low metallicity compared to other multiple-planet systems of giant stars, this system is an important sample to understand the influence of stellar metallicity on multiplicity of planets.
\subsection{Dynamical properties of the $\gamma$ Lib system}
The period ratio of $\gamma$ Lib b and c is close to 7:3 $(P_c/P_b = 2.33)$, which is a rare configuration among both giant stars and dwarf stars.
There are several multiple systems which are similar to $\gamma$ Lib in terms of an orbital period ratio of planets.
HD 181433 c and d have a period ratio of $P_d/P_c = 2.26\ (P_c=962.0\>{\rm d},\ P_d=2172\>{\rm d})$ \citep{Bouchy2009}.
Although the orbital period ratio is close to 11/5, 9/4 and 7/3, a dynamical stability test favored 5:2 MMR \citep{Campanella2011}.
HIP 65407 b and c have a period ratio of $P_c/P_b = 2.39\ (P_b=28.125\>{\rm d},\ P_c=67.30\>{\rm d})$ \citep{Hebrard2016} which is near 12:5 MMR.
However, a stability analysis revealed 12:5 MMR was unstable, and HIP 65407 b and c were found to be just outside this resonance \citep{Hebrard2016}.
47 UMa b and c have a period ratio of $P_c/P_b = 2.38\ (P_b=1089.0\>{\rm d},\ P_c=2594\>{\rm d})$ \citep{Fischer2002}, and this system was implied to participate in 7:3 MMR \citep{Laughlin2002}.
However, \citet{Wittenmyer2007} showed the orbital period of 47 UMa c obtained in \citet{Fischer2002} was not conclusive due to insufficient RV data.
According to \citet{Wittenmyer2007}, a longer orbital period of 47 UMa c $(P_c=7586\>{\rm d})$ was favored.

\begin{figure}
 \begin{minipage}{0.48\hsize}
\begin{center}
\includegraphics[width=0.85\textwidth]{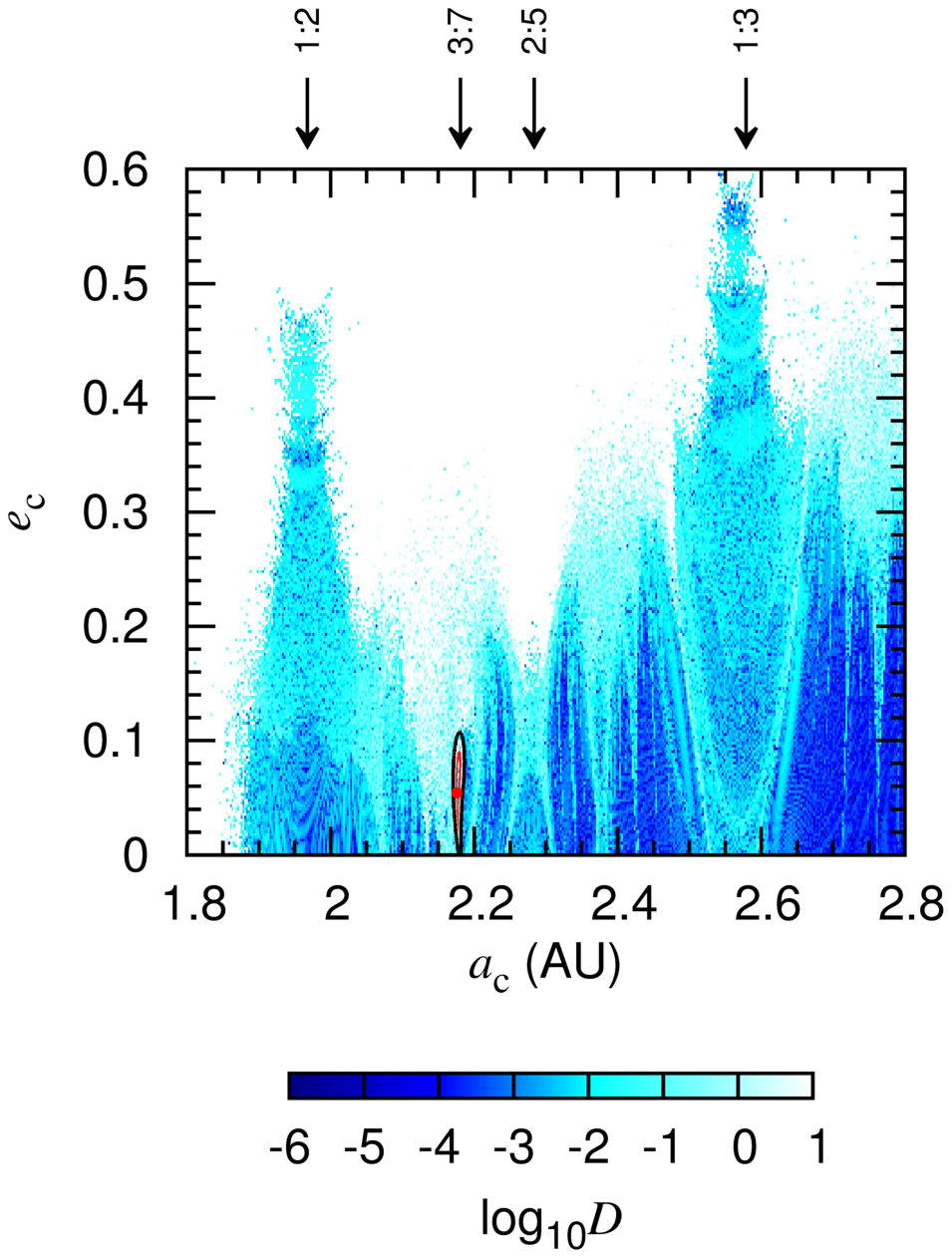}
\end{center}
\end{minipage}
 \hfill
 \begin{minipage}{0.48\hsize}
\begin{center}
\includegraphics[width=0.85\textwidth]{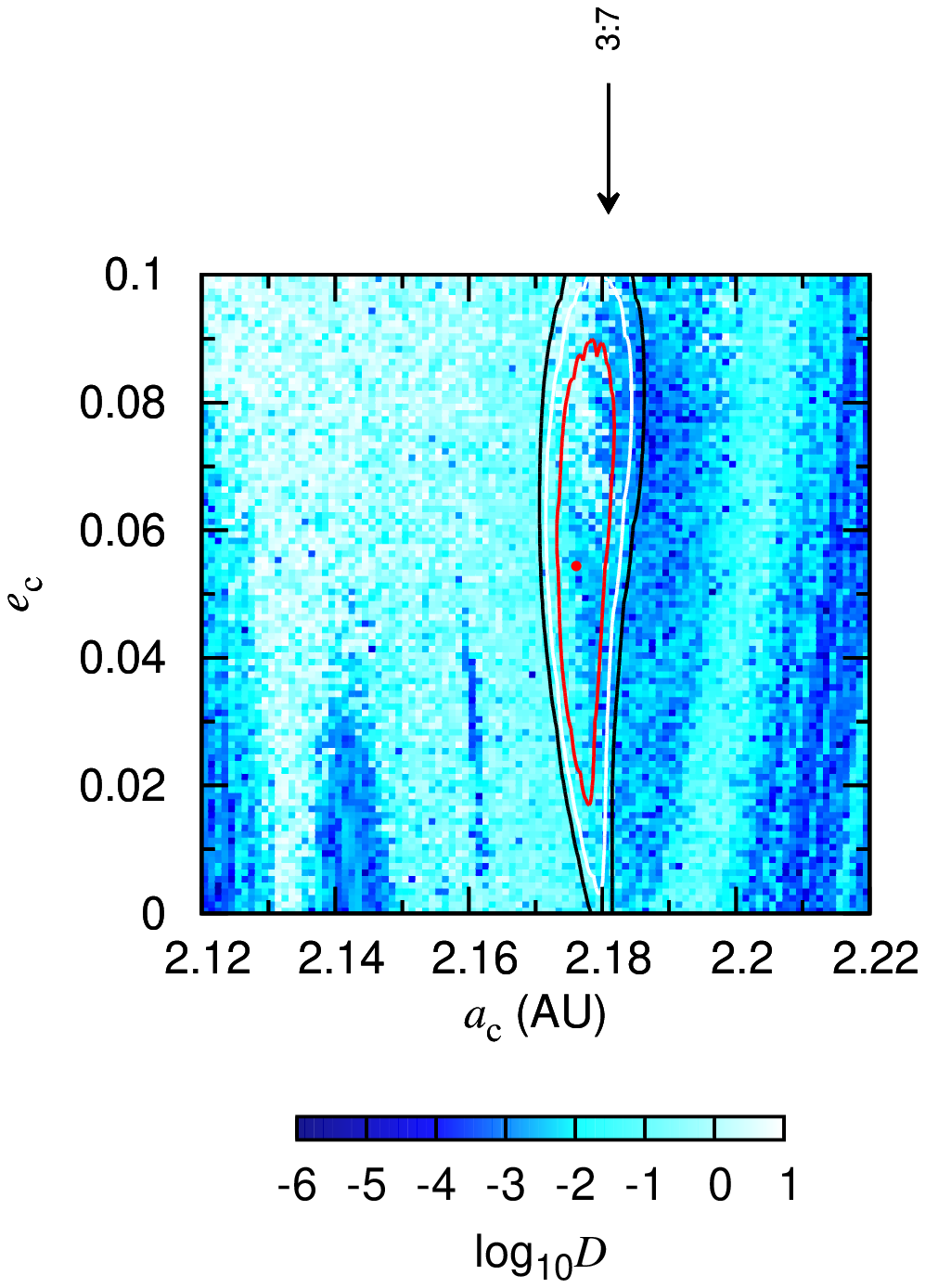}
\end{center}
\end{minipage}

\caption{Left: stability map in $(a, e)$ plane of the outer planet ($\gamma$ Lib c) with orbital inclination $i=90^{\circ}$.
The color scale shows the stability index $D$ (blue region is stable).
The red point represents the best-fitted value obtained with our dynamical analysis.
The red, white and black lines represent $1\,\sigma$, $2\,\sigma$ and $3\,\sigma$ errors of the best-fitted value, respectively.
The arrows indicate the values of $a_{\rm c}$ for MMRs with respect to the inner planet ($\gamma$ Lib b).
Right: enlarged view around the best-fitted values in the left panel.}
\label{fig:acec}
\end{figure}

\begin{figure}
\begin{center}
\includegraphics[width=0.6\textwidth]{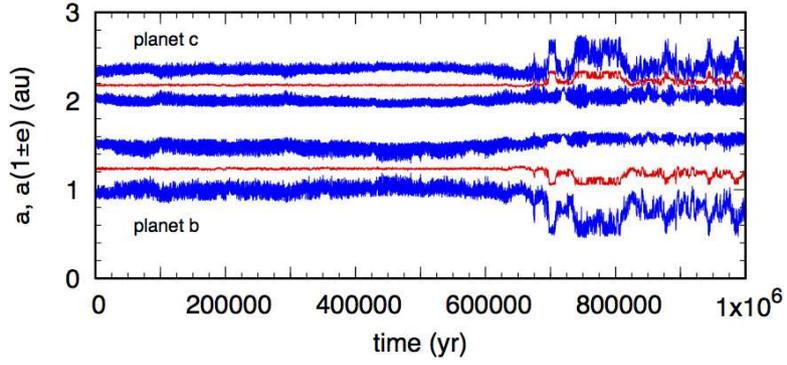}
\end{center}
\caption{Evolution of semimajor axis, $a$, pericenter distances, $a(1-e)$, and apocenter distances, $a(1+e)$ for $\gamma$ Lib b and c.
Red lines represent semimajor axis, and blue lines represent pericenter distances (lower ones for each planet) and apocenter distances (upper ones for each planet).}
\label{fig:1e6}
\end{figure}

In order to investigate the orbital properties of the $\gamma$ Lib system in more detail, we performed 3-body dynamical simulation for the system including gravitational interaction between planets (e.g. \cite{Sato2013b}, 2016). 
We here assumed that the planets are coplanar and are prograde. 
We used a fourth-order Hermite scheme for the numerical integration \citep{kokubo:1998}.
Figure~\ref{fig:acec} shows the map of the stability index $D=|\langle n_2 \rangle -\langle n_1 \rangle |$ of the system in $(a, e)$ plane of the planet c with orbital inclination $i=90^{\circ}$, 
where $\langle n_1 \rangle$ and $\langle n_2 \rangle$ are averages of mean motion of planet c obtained from 1000 Kepler periods ($\sim$ 3200$\>$yr) numerical integration and from the next 1000 Kepler periods integration, respectively (see \cite{couetdic:2010} for the details of the stability analysis).
Through comparison with the 5 times longer integrations, we found that the system is regular with about 95\% probability when $\log_{10} D \le -4$, and chaotic when $\log_{10} D > -2$.
The red point in the panel represents the best-fitted $a_c$ and $e_c$ to the observed RVs that are obtained in this dynamical analysis, and the red line represents their $1\,\sigma$ errors (Table~\ref{dtab2}). 
These values are consistent with those obtained from the double Keplerian model. 
We can see that the best-fitted pair of ($a_c$, $e_c$) is located at the edge of the stable area with slightly larger $a_c$.
Figure~\ref{fig:1e6} shows the evolution of semimajor axis and eccentricity for $\gamma$ Lib b and c, for which we adopt the best-fitted orbital parameters obtained in the dynamical analysis.
Indeed the system become unstable after $\sim$ 0.6 Myr.
Although we can still find stable configurations (dark-blue dots in figure~\ref{fig:acec}) within the $1\,\sigma$ of the best-fitted values, we can not totally say that the best-fitted orbits to the observed RV data is regular.
We also found that there are almost no stable area around $3\,\sigma$ of the best-fitted coplanar prograde orbits for $i\le 70^{\circ}$. 
Thus, we can say that the true masses of the planets deviate from the minimum masses by not more than $6$\%, which ensures that the system is a planetary system.

\begin{figure}
\begin{center}
\includegraphics[width=8cm]{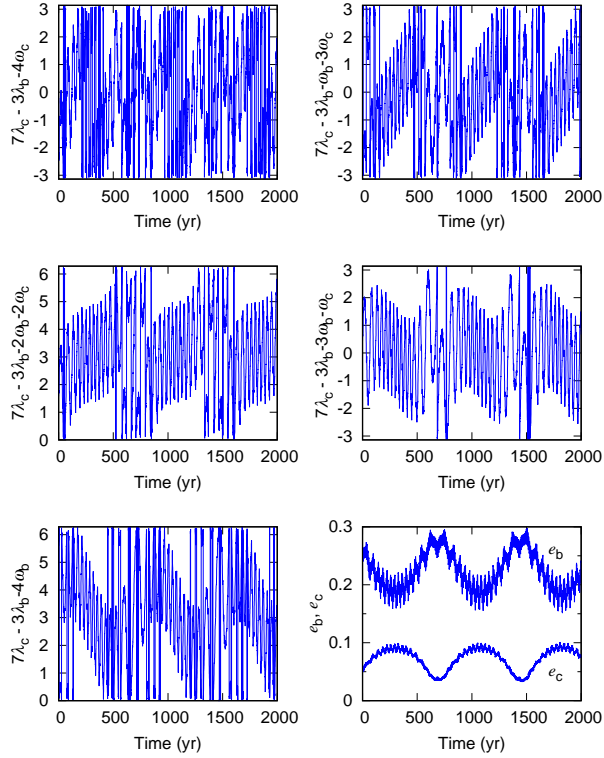}
\end{center}
\caption{Time evolution of the resonant angles (radian) and eccentricities for the best-fitted orbital parameters obtained with our dynamical analysis.}
\label{fig:libration1}
\end{figure}

The stable area with $a_c \simeq 2.18\>{\rm au}$ seen in Figure~\ref{fig:acec} is related to the 7:3 MMR for planet b and c. 
Since the best-fitted pair of $(a_c, e_c)$ is located at the edge of the resonance area, we investigated behaviors of related resonant angles. 
Figure~\ref{fig:libration1} shows time evolution of five possible resonant angles associated with the 7:3 commensurability in 2D case for the best-fitted orbital parameters obtained with our dynamical analysis. 
We found one resonant angle $\psi = 7\lambda_c - 3\lambda_b - 3\omega_b - \omega_c$ (right middle panel), 
where $\lambda$ and $\omega$ are mean longitude and longitude of pericenter, respectively, 
librates with a relatively large semi-amplitude of about 120$^{\circ}$ except when the eccentricity of the planet b approaches 0.3, while all of the other angles circulate.
We also found that the same resonant angle librates with a smaller semi-amplitude of about $70^{\circ}$ for longer integration in the case that $a_c$ is larger by $3\,\sigma$ from the best-fitted value. 
The resonant angle $7\lambda_c - 3\lambda_b - 4\omega_b$ (left bottom panel) also librates in this case, though the amplitude is rather large.
Combined with the result that the best-fitted orbits are not totally regular as shown in Figure~\ref{fig:acec}, we suggest that the true $a_c$ value is $2$--$3\,\sigma$ larger than the best-fitted value, and the planet b and c are in 7:3 MMR.

\bigskip
\begin{ack}
This research is based on data collected at the Okayama Astrophysical Observatory (OAO), which is operated by National Astronomical Observatory of Japan. 
We are grateful to all the staff members of OAO for their support during the observations.
Data of HD 113226 were partially obtained during an engineering time at OAO. 
We thank the observatory for allowing us to use the data obtained during that time.
We thank students of Tokyo Institute of Technology and Kobe University for their kind help with the observations at OAO. 
B.S. was partially supported by MEXTʼs program ``Promotion of Environmental Improvement for Independence of Young Researchers" under the Special Coordination Funds for Promoting Science and Technology, and by Grant-in-Aid for Young Scientists (B) 17740106 and 20740101 and Grant-in-Aid for Scientific Research (C) 23540263 from the Japan Society for the Promotion of Science (JSPS). 
H.I. is supported by Grant-in-Aid for Scientific Research (A) 23244038 from JSPS. 
This research has made use of the SIMBAD database, operated at CDS, Strasbourg, France.
\end{ack}

\appendix
\section*{Effect of IP variability on line profile analysis} \label{appendix}
The spectral line-profile deformation could be caused by IP variations as well as stellar surface modulation.
Since 24 Boo and $\gamma$ Lib have rotational velocities which are comparable to or lower than the velocity resolution of HIDES, BIS of CCF (BIS$_{\rm CCF}$) may be affected by IP variations.
To evaluate the effect of IP variation on line-profile analysis, we performed the analysis for a chromospherically inactive G-type giant, HD 113226.
The star has a rotational velocity of $v\sin{i}=3.05\>{\rm km\>s^{-1}}$ \citep{Takeda2008} and shows small RV variation.
We calculate BIS of mean IP (BIS$_{\rm IP}$) that is derived from the RV analysis (section \ref{sec:rv}). 
Although the IP is estimated by using the different wavelength region from that for the bisector analysis, we can expect that the variability is basically the same.

\begin{figure}
\begin{center}
\includegraphics[width=10cm]{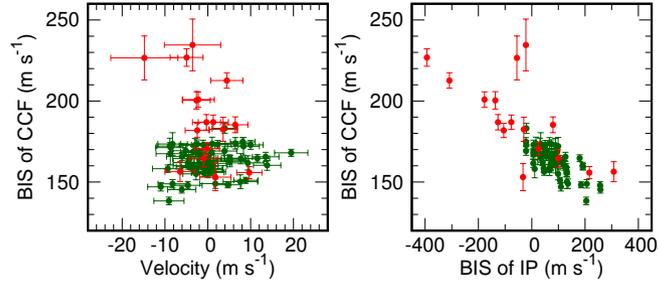}
\end{center}
\caption{Left: BIS$_{\rm CCF}$ plotted as a function of the observed RV for HD 113226. Red and green circles are data taken with slit- and fiber-mode, respectively. Right: BIS$_{\rm CCF}$ plotted as a function of BIS$_{\rm IP}$.}
\label{rv_bis_ip_113226}
\end{figure}

Figure~\ref{rv_bis_ip_113226} shows a relation between BIS$_{\rm CCF}$ and RV (left panel), and BIS$_{\rm CCF}$ and BIS$_{\rm IP}$ (right panel) for HD 113226.
While there is no correlation between BIS$_{\rm CCF}$ and RV, we can see strong correlation between BIS$_{\rm CCF}$ and BIS$_{\rm IP}$ ($r=-0.67$ for slit-spectrum and $r=-0.71$ for fiber-spectrum).
Considering the small RV variation and small rotational velocity, BIS$_{\rm CCF}$ of HD 113226 should reflect the extent of line-profile deformation caused by IP variation.
Therefore the RMSs of BIS$_{\rm CCF}$, $26.2\>{\rm m\>s^{-1}}$ for slit-spectrum and $8.8\>{\rm m\>s^{-1}}$ for fiber-spectrum, can be regarded as a typical line-profile deformation caused by IP variations of HIDES.

\begin{figure}
\begin{center}
\includegraphics[width=10cm]{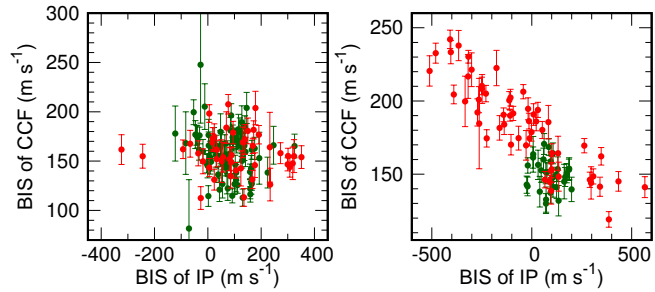}
\end{center}
\caption{Left: BIS$_{\rm CCF}$ plotted as a function of BIS$_{\rm IP}$ for 24 Boo. Red and green circles are data taken with slit- and fiber-mode, respectively. Right: BIS$_{\rm CCF}$ plotted as a function of BIS$_{\rm IP}$ for $\gamma$ Lib.}
\label{rv_bis_ip_127_138}
\end{figure}

Figure~\ref{rv_bis_ip_127_138} shows a relation between BIS$_{\rm CCF}$ and BIS$_{\rm IP}$ for 24 Boo (left) and $\gamma$ Lib (right).
In the case of 24 Boo, we can see no significant correlation between BIS$_{\rm CCF}$ and BIS$_{\rm IP}$.
Furthermore, the RMSs of BIS$_{\rm CCF}$ are almost the same between two modes ($18.5\>{\rm m\>s^{-1}}$ and $22.3\>{\rm m\>s^{-1}}$ for slit-spectra and fiber-spectra, respectively).
This implies that stellar surface modulation is a dominant factor in BIS$_{\rm CCF}$ variations of 24 Boo rather than the IP variability.
As for $\gamma$ Lib, BIS$_{\rm CCF}$ of slit-spectra is strongly correlated with BIS$_{\rm IP}$ ($r=-0.86$) and the RMSs of BIS$_{\rm CCF}$ are quite different in two modes ($29.2\>{\rm m\>s^{-1}}$ for slit-spectrum and $10.4\>{\rm m\>s^{-1}}$ for fiber-spectrum).
Therefore the IP variability is a dominant factor in the BIS$_{\rm CCF}$ variations for $\gamma$ Lib and the stellar surface modulation is not significant.

\end{document}